\documentstyle[12pt]{article}
\def\PRL #1 #2 #3{{\sl Phys. Rev. Lett.} {\bf#1} (#2) #3}
\def\NPB #1 #2 #3{{\sl Nucl. Phys.} {\bf B#1} (#2) #3}
\def\NPBFS #1 #2 #3 #4{{\sl Nucl. Phys.} {\bf B#2} [FS#1] (#3) #4}
\def\CMP #1 #2 #3{{\sl Commun. Math. Phys.} {\bf #1} (#2) #3}
\def\PRD #1 #2 #3{{\sl Phys. Rev.} {\bf D#1} (#2) #3}
\def\PLA #1 #2 #3{{\sl Phys. Lett.} {\bf #1A} (#2) #3}
\def\PLB #1 #2 #3{{\sl Phys. Lett.} {\bf #1B} (#2) #3}
\def\JMP #1 #2 #3{{\sl J. Math. Phys.} {\bf #1} (#2) #3}
\def\PTP #1 #2 #3{{\sl Prog. Theor. Phys.} {\bf #1} (#2) #3}
\def\SPTP #1 #2 #3{{\sl Suppl. Prog. Theor. Phys.} {\bf #1} (#2) #3}
\def\AoP #1 #2 #3{{\sl Ann. of Phys.} {\bf #1} (#2) #3}
\def\PNAS #1 #2 #3{{\sl Proc. Natl. Acad. Sci. USA} {\bf #1} (#2) #3}
\def\RMP #1 #2 #3{{\sl Rev. Mod. Phys.} {\bf #1} (#2) #3}
\def\PR #1 #2 #3{{\sl Phys. Reports} {\bf #1} (#2) #3}
\def\AoM #1 #2 #3{{\sl Ann. of Math.} {\bf #1} (#2) #3}
\def\UMN #1 #2 #3{{\sl Usp. Mat. Nauk} {\bf #1} (#2) #3}
\def\FAP #1 #2 #3{{\sl Funkt. Anal. Prilozheniya} {\bf #1} (#2) #3}
\def\FAaIA #1 #2 #3{{\sl Functional Analysis and Its Application} {\bf
#1} (#2) #3}
\def\BAMS #1 #2 #3{{\sl Bull. Am. Math. Soc.} {\bf #1} (#2)
#3} \def\TAMS #1 #2 #3{{\sl Trans. Am. Math. Soc.} {\bf #1} (#2) #3}
\def\InvM #1 #2 #3{{\sl Invent. Math.} {\bf #1} (#2) #3}
\def\LMP #1 #2 #3{{\sl Letters in Math. Phys.} {\bf #1} (#2) #3}
\def\IJMPA #1 #2 #3{{\sl Int. J. Mod. Phys.} {\bf A#1} (#2) #3}
\def\AdM #1 #2 #3{{\sl Advances in Math.} {\bf #1} (#2) #3}
\def\RMaP #1 #2 #3{{\sl Reports on Math. Phys.} {\bf #1} (#2) #3}
\def\IJM #1 #2 #3{{\sl Ill. J. Math.} {\bf #1} (#2) #3}
\def\APP #1 #2 #3{{\sl Acta Phys. Polon.} {\bf #1} (#2) #3}
\def\TMP #1 #2 #3{{\sl Theor. Mat. Phys.} {\bf #1} (#2) #3}
\def\JPA #1 #2 #3{{\sl J. Physics} {\bf A#1} (#2) #3}
\def\JSM #1 #2 #3{{\sl J. Soviet Math.} {\bf #1} (#2) #3}
\def\MPLA #1 #2 #3{{\sl Mod. Phys. Lett.} {\bf A#1} (#2) #3}
\def\JETP #1 #2 #3{{\sl Sov. Phys. JETP} {\bf #1} (#2) #3}
\def\JETPL #1 #2 #3{{\sl  Sov. Phys. JETP Lett.} {\bf #1} (#2) #3}
\def\PHSA #1 #2 #3{{\sl Physica} {\bf A#1} (#2) #3}
\def\CQG #1 #2 #3{{\sl Class. Quantum Grav.} {\bf #1} (#2) #3}
\def\SJNP #1 #2 #3{{\sl Sov. J. Nucl. Phys. (Yadern.Fiz.)} {\bf #1} (#2)
#3}
\def\d{\delta}\def\eps{\epsilon}

\def\L{\Lambda}

\def\cf{\cal F}
\def\be{\begin{equation}}\def\ee{\end{equation}}

\newcommand{\p}[1]{(\ref{#1})}
\begin{document}
\thispagestyle{empty}
\renewcommand{\thefootnote}{\fnsymbol{footnote}}
\begin{flushright}
HUB-EP-98/47\\
hep-th/9808049\\
\end{flushright}

\vspace{1truecm}
\begin{center}
{\large\bf Dual Actions for Chiral Bosons}\footnote{
Talk given at the International Workshop ``Supersymmetry and Integrable
Systems" (June 22--26,1998, Dubna, Russia) and at the Xth
School--Seminar
on Recent Problems in Theoretical and Mathematical Physics ``VOLGA
10'98"
(June 22--July 2, 1998, Kazan, Russia).
}

\vspace{0.4cm} Alexey Maznytsia$^1$, Christian R. Preitschopf$^2$
and
Dmitri Sorokin$^2$
\footnote{Alexander von Humboldt Fellow \\
\phantom{mn} On leave from Kharkov Institute of Physics and
Technology, Kharkov, 310108, Ukraine}

\vspace{0.5cm}
$^1${\it Department of Physics and Technology,
Kharkov State University \\
310108, Kharkov, Ukraine\\
e--mail: alex$\_{}$maznytsia@hotmail.com
}

\vspace{0.5cm}
$^2${\it Humboldt-Universit\"at zu Berlin\\
Institut f\"ur Physik\\
Invalidenstrasse 110, D-10115 Berlin, Germany\\
e--mails: preitsch,sorokin@physik.hu-berlin.de}

\end{center}

\vspace{0.5cm}
\vspace{0.3cm}

\begin{center}
{\bf Abstract}
\end{center}

\vspace{0.3cm}

We study duality properties of
actions for chiral boson fields in various space--time dimensions using
$D=2$ and $D=6$ cases as examples. As a result we get dual
covariant formulations of chiral bosons.

\renewcommand{\thefootnote}{\arabic{footnote}}
\setcounter{footnote}0
%\newpage

\renewcommand{\thefootnote}{\arabic{footnote}}
\setcounter{footnote}0

\section{Introduction}

During the last few years the study of duality--symmetric and self--dual
fields (or chiral bosons) attracted much attention because of the role
they play in theoretical models related to superstring
theories revealing various types of dualities
of these theories. The fields of this
kind appear on the worldvolumes of  heterotic strings, M--theory
five--branes and in several field--theoretical
limits of M--theory and superstrings.

Chiral bosons are associated with differential $p$-forms
$A^{(p)}$ in the $D=2(p+1)$--dimensional space--time, whose
external differential
$F^{(p+1)}(A)=dA^{(p)}$ is restricted by a self--duality
condition
\be \label{form}
{\cal F}^{(p+1)}\equiv F^{(p+1)}(A)-^*F^{(p+1)}(A)=0;\quad
{\cal F}^{(p+1)}=-^*{\cal F}^{(p+1)}\quad.
\ee
When the dimension of a Lorentzian space--time
\footnote{In what follows we use a metric of mostly positive
signature.}
is twice odd the chiral form is real,
while if it is twice even the chiral field is complex (or described by
two real p--forms). For instance,
in the $D=4$ case ($p=1$) a field theory with two vector
potentials whose field strengths are connected by the duality
relation \p{form} describes duality--symmetric Maxwell electrodynamics
\cite{zwanziger}--\cite{pstch1}.

The first--order differential equation \p{form} is an equation of motion
which defines the dynamics of
the chiral boson. This feature distinguishes chiral bosons from
other bosonic fields whose equations of motion are usually
second--order differential equations. Thus it seems natural to try to
construct an action for chiral bosons in a first order form
\cite{infinite1,infinite2}.  In such a formulation eq. \p{form} appears as
the equation of motion of an auxiliary Lagrange multiplier field. It turns
out that for the Lagrange multiplier itself not to carry propagating
degrees of freedom one has to introduce an infinite number of auxiliary
fields ``compensating" the dynamics of each other.

An action of another kind (which will be the subject of our discussion)
\cite{pst,pstch1,pstch2} is quadratic in field strengths and contains
only
one auxiliary scalar field ensuring manifest Lorentz invariance of the
construction.
This formulation is a covariant generalization
of non manifestly space-time invariant actions for chiral bosons
\cite{zwanziger,deser,ss,fj,henn}
and it has been
used for the  construction of the effective action for the
M--theory super--five--brane \cite{5br} coupled to a duality--symmetric
$D=11$ supergravity \cite{bbs}, Type IIB
supergravity \cite{2bsg} and for some other models \cite{6dn2}.

In this paper we discuss duality properties of the covariant chiral
boson action
in diverse space--time dimensions considering the $D=2$ and $D=6$ cases as
examples. The results obtained for the six--dimensional chiral
boson represent basic duality features inherent to this formulation in
any
space--time dimension \cite{mps}.

\section{Doubly Self--Dual Action In $D=2$}

We begin with the simplest two--dimensional free chiral boson model.
The literature devoted to studying $D=2$ chiral bosons and their
quantization is very extensive, and we are able to refer the reader
only to some of the papers where different approaches were addressed
\cite{fj,infinite1}, \cite{si}--\cite{gates}.

In the $D=2$ case the chiral boson field is a scalar whose
``field strength" $F_m(\phi )={\partial}_m\phi $ satisfies the following
on--shell self--duality condition
\be \label{2sd}
{\cf}_m(\phi )={\partial}_m\phi -{\eps}_{mn}{\partial}^n\phi=0,\quad
{\cf}_m(\phi )=-{\eps}_{mn}{\cf}^n(\phi)
\ee
This condition can be obtained from the
action \cite{pstch1,pstch2}
\be \label{2pst}
S=\int d^2x[-{1\over 2}F_m(\phi )F^m(\phi )+
{1\over{2({\partial}_ra)({\partial}^ra)}}
({\partial}^ma{\cf}_m(\phi ))^2],
\ee
where $a(x)$ is the auxiliary scalar field mentioned above. It enters
the
action in a non--polynomial way, and to avoid singularities,
we require $({\partial}_ra)({\partial}^ra)\neq 0$. This condition
reflects a non--trivial topological structure of this theory and is
present in the formulations of this kind in any space-time dimension.

The action \p{2pst} possesses the following set of local symmetries
\cite{pstch1,pstch2,cher}
\be \label{2asymm}
\d a=\varphi(x) ,\qquad
\d \phi ={{\varphi}\over{({\partial}a)^2}}
{\cf}^m(\phi ){\partial}_ma,
\ee
\be \label{2fsymm}
\d \phi =f(a(x)),
\ee
with the parameters ${\varphi}(x)$ and $f(a)$, respectively. Note that
the
latter depends on $x$ only through the field $a(x)$ and there is no
any first class constraint associated with this symmetry.
The only first class constraint, which one finds in the
Hamiltonian formulation of this model, generates the symmetry \p{2asymm}
\cite{pstch2}. The self--duality condition \p{2sd} appears as a
general solution to the $\phi$--field equations of motion,
obtained from the action \p{2pst}, upon gauge fixing the symmetries
\p{2asymm} and \p{2fsymm}.

    Now let us study the duality properties of this action. There
are two fields which can be dualized, $\phi(x)$ and $a(x)$.

To get the $\phi $--dual formulation, we replace \p{2pst} with
\be \label{2fparent}
S=\int d^2x[-{1\over 2}F_mF^m+
{1\over{2({\partial}_ra)({\partial}^ra)}}
({\partial}^ma{\cf}_m)^2+G^m(F_m-{\partial}_m{\phi})].
\ee
In this action $F_m$ and $G_m$ are regarded as independent vector fields
and ${\cf}_m\equiv F_m-{\eps}_{mn}F^n$. The classical equivalence of the
actions \p{2pst} and \p{2fparent} is evident. Varying \p{2fparent} with
respect to the Lagrange multiplier $G_m$ we obtain
$$
F_m=\partial_m\phi,
$$
which yields \p{2pst} when substituted into  \p{2fparent}.
The variation of the  action \p{2fparent} with respect to $F_m$
regarded as another
auxiliary field  produces an expression for $G_m$ in
terms of $F_m$ which, when substituted into \p{2pst}, results in a
dual action. If we perform this procedure, we shall realize that the
dual
action obtained this way coincides with \p{2pst}, so the action \p{2pst}
is self--dual with respect to the dualization of the chiral boson field.
Note that this situation happens for the  free chiral boson model in
any even space--time dimension and reflects the basic (self--duality)
property of the chiral bosons. The
reader may find an explicit proof of this self--duality of $D=2$ and
$D=4$ chiral boson actions in \cite{mps}.

Consider now the properties of \p{2pst} with respect to the dualization of
the auxiliary scalar $a(x)$. In order to do this we replace \p{2pst}
with the
following classically equivalent action
\be \label{2parenta}
S=\int d^2x[-{1\over 2}F_m(\phi )F^m(\phi )+
{1\over{2u^ru_r}}
(u^m{\cf}_m(\phi ))^2+v^m(u_m-{\partial}_ma)]
\ee
which contains the independent auxiliary fields $v^m$ and $u^m$. The
equation of motion for the field $v^m$  is
\be\label{u}
u_m={\partial}_ma.
\ee
It reduces the model to the one described by \p{2pst}.

The variation of \p{2parenta} with respect to $u^m$ produces the
constraint
\be \label{2vconstr}
v^m={1\over{(u^ru_r)^2}}{\eps}^{mn}u_n(u^p{\cf}_p(\phi ))^2,
\ee
from which it follows, in particular, that the normalized vectors $u^m$
and $v^m$ are dual to each other
\footnote{Note that this relation holds only off the mass shell,
i.e. when ${\cal F}_m$ is non--zero and the relation \p{2vconstr} is
non--degenerate.}:
\be \label{2conseq}
{u^m\over{\sqrt{(u)^2}}}=
\eps^{mn}{v_n\over{\sqrt{-(v)^2}}}.
\ee

Substituting \p{2vconstr} into \p{2parenta}, we get
\be \label{2dualv}
S=\int d^2x[-{1\over 2}F_m(\phi )F^m(\phi )-
{1\over{2v^rv_r}}
(v^m{\cf}_m(\phi ))^2+a({\partial}_mv^m)].
\ee
The equation of motion of the field $a(x)$ allows us to express $v^m$ as
\be \label{2curl}
v^m={\eps}^{mn}{\partial}_nb,
\ee
where $b(x)$ is a scalar field.
Inserting this expression back into the action \p{2dualv} and taking
into
account the anti--self--duality of ${\cal F}^m(\phi )$, we recover the
action \p{2pst} with the field $a(x)$ replaced by $b(x)$. Thus, the
action
\p{2pst} describing the free two--dimensional chiral scalar is
self--dual with respect to the dualization of the auxiliary scalar field
$a(x)$ as well.

\section{Dual Actions in $D=6$}

The situation changes when we consider such a dualization in
a space--time of higher dimension.
Double self--duality which we observed in $D=2$ is due to the duality
relation between scalars \p{u}, \p{2conseq} and  \p{2curl}, which hold
only in
$D=2$. In higher space--time dimensions scalars are dual to tensors
of a rank $D-2>0$ and therefore the field contents of the dual actions
will have no chance to coincide anymore. In this Section we demonstrate
this fact with the example of a free chiral boson in $D=6$.

A six--dimensional chiral boson field is a real antisymmetric field
$A_{mn}$, $(m,n=0,...,5)$ whose field strength
\be \label{6stress}
F_{mnp}(A)={\partial}_mA_{np}+{\partial}_nA_{pm}+{\partial}_pA_{mn}
\ee
is restricted (on the mass shell) by the self--duality condition
\be \label{6sd}
{\cf}_{mnp}(A)={F}_{mnp}(A)-{1\over{3!}}{\eps}_{mnpqrs}F^{qrs}(A)=0,
\quad
{\cf}_{mnp}(A)=-{1\over{3!}}{\eps}_{mnpqrs}{\cf}^{qrs}(A).
\ee
A $D=6$ analogue of the action \p{2pst} is \cite{pstch2}
\be \label{6pst}
S=\int d^6x[-{1\over 6}F_{mnp}(A)F^{mnp}(A)+
{1\over{2({\partial}_ra)({\partial}^ra)}}
{\partial}^ma{\cf}_{mnp}(A){\cf}^{npq}(A){\partial}_qa],
\ee
This action is invariant with respect to the following local
transformations of the fields $a$ and $A^{mn}$ \cite{pstch2}:
\be \label{6asymm}
\d a=\varphi(x) ,\qquad
\d A_{mn} ={{\varphi}\over{2({\partial}a)^2}}
{\cf}_{mnp}(A){\partial}^pa;
\ee
\be \label{6Asymm1}
\d A_{mn}={\partial}_{[m}{\Phi}_{n]};
\ee
\be \label{6Asymm2}
\d A_{mn}={\Psi}_{[m}{\partial}_{n]}a.
\ee
Note that in contrast to the two--dimensional model
\p{2pst} all the symmetries \p{6asymm}--\p{6Asymm2} are full--fledged
local
symmetries (i.e. in the Hamiltonian formulation of the
model there is a first class constraint associated with each of these
symmetries).
Using these symmetries, we can reduce the general solution of the
equations of motion of $A_{mn}$ derived from \p{6pst}
to the self--duality condition \p{6sd}.

Consider now the duality properties of the action \p{6pst}
with respect to the duality transform of the field $a(x)$.
In order to do this (as in the previous Section) we replace \p{6pst}
by
\be \label{6parent}
S=\int d^6x[-{1\over 6}F_{mnp}(A)F^{mnp}(A)+
{1\over{2u^ru_r}}u^m{\cf}_{mnp}(A){\cf}^{npq}(A)u_q+v^m(u_m-{\partial}_ma)],
\ee
which is equivalent to \p{6pst} as a consequence of the equation of
motion of the Lagrange multiplier $v^m$. The variation of this action
with respect to the field $u^m$ gives an expression for $v^m$ in terms
of other fields
\be \label{v6constr}
v^m=-{1\over{u^2}}{\cf}^{mnp}(A){\cf}_{npq}(A)u^q+{1\over{(u^2)^2}}
u^mu^n{\cf}_{npq}(A){\cf}^{pqr}(A)u_r,
\ee
or, because of the anti--self--duality of ${\cf}^{mnp}(A)$
\be\label{v6}
v^m={1\over{2(u^2)^2}}\varepsilon^{mlnpqr}u_l{\cf}_{nps}u^s{\cf}_{qrt}u^t
\ee
from which it follows that, in particular, the vectors $u^m$ and
$v^m$ are orthogonal
\be \label{v6conseq}
u^mv_m=0,
\ee
and
\be \label{v6conseq2}
{1\over{2u^ru_r}}u^m{\cf}_{mnp}(A){\cf}^{npq}(A)u_q=
-{1\over{2v^rv_r}}v^m{\cf}_{mnp}(A){\cf}^{npq}(A)v_q+
2{\tilde{\cal F}}^m{\tilde{\cal F}}_m,
\ee
where
\be\label{fm}
{\tilde{\cal F}}_m\equiv {1\over{\sqrt{u^2v^2}}}{\cal F}_{mnp}v^nu^p.
\ee
Taking into account eq. \p{v6constr} and the self--duality of
${\cal F}^{mnp}(A)$, one can show that the second term on the r.h.s. of
\p{v6conseq2} vanishes identically. To see this, replace one of $v^n$
in ${\tilde{\cal F}}^m{\tilde{\cal F}}_m$ with its expression \p{v6},
then
\be\label{fmfm}
{\tilde{\cal F}}^m{\tilde{\cal F}}_m=
{1\over{2(u^2)^2\sqrt{u^2v^2}}}{\tilde{\cal F}}_m{\cf}^{mnp}u_p
\varepsilon_{ntr_1r_2s_1s_2}u^l{\cf}^{r_1r_2s}u_s{\cf}^{s_1s_2t}u_t.
\ee
Now replace the second ${\cf}^{mnp}$ on the right hand side of \p{fmfm}
with its antiselfdual and ``eliminate" two epsilon--tensors. Eq.
\p{fmfm} takes the form
\be\label{fmfm1}
{\tilde{\cal F}}^m{\tilde{\cal F}}_m=
{2\over{(u^2)\sqrt{u^2v^2}}}{\tilde{\cal F}}_m{\cf}^{mnp}u_p
{\cf}^{rst}u_t{\cf}_{nrs}.
\ee
In virtue of \p{v6constr} the last two ${\cf}$ in \p{fmfm1}
can be replaced with $v^m$ which results in
$$
{\tilde{\cal F}}^m{\tilde{\cal F}}_m=-2{\tilde{\cal F}}^m{\tilde{\cal
F}}_m, ~~~\Rightarrow~~~ {\tilde{\cal F}}^m{\tilde{\cal F}}_m=0.
$$
Substituting \p{v6conseq2} with ${\tilde{\cal F}}^m{\tilde{\cal F}}_m=0$
into the action \p{6parent}, we get
\be \label{6inter}
S=\int d^6x[-{1\over 6}F_{mnp}(A)F^{mnp}(A)-
{1\over{2v^rv_r}}v^m{\cf}_{mnp}(A){\cf}^{npq}(A)v_q+a{\partial}_mv^m].
\ee
Then, solving the dynamical equation of the field $a(x)$ we express
$v^m(x)$ in terms of a 4-form field $B_{mnpq}$:
$$
v^m={\eps}^{mnpqrs}{\partial}_nB_{pqrs},
$$
and substituting this expression into \p{6inter} we obtain the dual
action for the $D=6$ chiral boson in the following form
\be \label{6dual}
S=\int d^6x[-{1\over 6}F_{mnp}(A)F^{mnp}(A)-
{1\over{2v^rv_r}}v^m{\cf}_{mnp}(A){\cf}^{npq}(A)v_q],
\ee
$$
v^m={\eps}^{mnpqrs}{\partial}_nB_{pqrs}.
$$
We see that, as in the $D=2$ case, the vector fields $u^m$ and $v^m$
are dual to each other in a sense that $u^m$ is the ``field strength" of
the scalar field $a(x)$ and $v^m$ is the dual ``field strength"
of the
4-form field $B_{mnpq}$. Therefore, as we expected, the dual action
\p{6dual} does not coincide with the initial one. Moreover it has a
different symmetry structure. Eq. \p{6dual} is invariant under the
following local transformations
$$
\d B_{mnpq}={\partial}_{[m}C_{npq]};
$$
\be \label{Bsymm}
\d B_{mnpq}={\eps}_{mnpqrs}v^r{\L}^s,\quad
\d A_{mn}=2{\eps}_{mnpqrs}{\cf}^{pqt}v_tv^r{\L}^s,
\ee
$$
\d
v^m={\L}^n{\partial}_nv^m-v^n{\partial}_n{\L}^m+v^m{\partial}_n{\L}^n,
$$
($C_{mnp}$ and ${\L}_m$ are the parameters), and under
\be \label{semiloc}
\d A_{mn}={1\over{\sqrt{v^2}}}{\eps}_{mnpqrs}v^p{\Phi}^{qrs},
\ee
where the parameter ${\Phi}^{mnp}$ is restricted by the differential
condition
\be \label{restr}
v^t{\partial}_t[{1\over{\sqrt{v^2}}}{\eps}_{mnpqrs}v^p{\Phi}^{qrs}]-
({\partial}_{[m}v^t){1\over{\sqrt{v^2}}}{\eps}_{n]tpqrs}v^p{\Phi}^{qrs}=0.
\ee
Because of this restriction eq. \p{semiloc} is not a conventional
local symmetry and
is an analogue of the symmetry \p{2fsymm} of the chiral scalar action
\p{2pst}. In order to see this consider \p{semiloc} in the
gauge
\be \label{gauge}
{v^m\over{\sqrt{v^2}}}=n^m,\quad n^m=const,
\ee
which we can impose using the symmetry \p{Bsymm}. Now eq.
\p{restr} takes the following form
$$
n^t{\partial}_t[{\eps}_{mnpqrs}n^p{\Phi}^{qrs}(x)]=0,
$$
which is solved in terms of an arbitrary 3--form depending on five
independent arguments
$$
{\Phi}_{mnp}={\Phi}_{mnp}(y), \quad
y^m=x^m-n^m(n_px^p),
$$
which are coordinates transversal to the
vector $n^m$. If we consider the symmetry \p{2fsymm} in the
non--covariant
gauge $a=n^m{\eps}_{mn}x^n$, we will see that its parameter also depends
on the coordinates transversal to the vector $n^m$.

Using the symmetries \p{Bsymm} and \p{semiloc}
one can reduce the equations of motion of $A_{mn}$
\be \label{nceq}
{\partial}_m({1\over{v^2}}v^{[m}{\cal F}^{np]q}(A)v_q)=0,
\ee
which follow from
\p{6dual}, to the self--duality
condition \p{6sd}. Thus, the dual actions \p{6pst} and \p{6dual} both
give a consistent description of the six--dimensional free chiral boson
field.

It is instructive to note that in the gauge \p{gauge}
the equations of motion \p{nceq} can be reduced to Maxwell--type
equations for a self--dual 3--form field strength. To see this notice
that the following identity holds for any $F^{(3)}=dA^{(2)}$
$$
{\partial}^m{\eps}_{mnpqrs}n^qF^{rst}(A)n_t=
-{1\over 3}(n^m{\partial}_m){\eps}_{npqrst}n^qF^{rst}(A).
$$
Using this identity one can rewrite eqs. \p{nceq} (with $v^m$ satisfying
\p{gauge}) in the form
$$
{\partial}_m{\cal T}^{mnp}=0,
$$
where
$$
{\cal T}^{mnp}=3n^{[m}F^{np]q}(A)n_q+
{1\over 2}{\eps}^{mnpqrs}n_qF_{rst}(A)n^t\equiv
{1\over 6}{\eps}^{mnpqrs}{\cal T}_{qrs}
$$
is a self--dual combination of $F_{mnp}$.
We observe that the corresponding anti--self--dual combination of
$F_{mnp}$ does not enter the $A_{mn}$--field equations of motion and,
hence, decouples from the classical degrees of freedom. This implies
from a somewhat different point of view that the model under consideration
indeed has the required properties to reproduce the self--duality
condition \p{6sd} and describes the dynamics of a single chiral boson.

\section{Dual Actions With External Sources}

In this Section we discuss the problem of coupling chiral
boson fields to external field sources. As in the previous Section
we will use the $D=6$ case as an example \cite{deser_s}, \cite{bm}.

If the $D=6$ gauge field ${A}_{mn}$ were not self--dual the action
\p{6pst} would not contain the second term, and its minimal coupling
to external field sources would be described solely by the standard term
\be \label{int}
S_{int}=-\int d^6x{j}^{mn}{A}_{mn},
\ee
where ${j}^{mn}$ is a conserved charged current
(${\partial}_m{j}^{mn}=0$). The conservation of the current ensures the
gauge symmetry of the action under $A_{mn}\rightarrow
{A}_{mn}+{\partial}_{[m}\varphi_{n]}$.

In the case of the chiral field the situation with coupling becomes much
more complicated. The interaction terms must be now compatible with the
symmetry structure \p{6asymm}--\p{6Asymm2} of the free self--dual
theory.
To find this term notice that, because of self--duality, the equation of
motion of the chiral field coincides with the Bianchi identities for
its field strength. Therefore,
in order to introduce sources
while maintaining the self--duality condition \p{6sd}, one should
require them to possess equal ``electric" and ``magnetic" charges, that
is to be dyons. This leads to a modification of the Bianchi identities
which acquire non--zero right hand side. This means that the field
strength is not simply the curl of the gauge field potential anymore and
includes a so called ``Dirac membrane" (a two--dimensional analogue of
the Dirac string \cite{dirac}) which accumulates on its surface
singularities of the chiral gauge fields associated with the charged
sources. The Dirac membrane is described by an antisymmetric tensor
$S_{mnp}$ which, by definition, satisfies the equation
\be \label{Seq}
{\partial}_mS^{mnp}=j^{np}.
\ee
Thus, in general, $S_{mnp}$ is a nonlocal solution of
\p{Seq} in terms of the current.

The chiral gauge field strength now gets modified
as follows
\be \label{mod}
F_{mnp}(A)\rightarrow {\hat F}_{mnp}\equiv F_{mnp}(A)+{1\over 6}
{\eps}_{mnpqrs}S^{qrs},
\ee
and the action describing the coupling of the chiral boson to charged
sources has the form \cite{bm}
\be \label{6psts}
S=\int d^6x[-{1\over 6}{\hat F}_{mnp}(A){\hat F}^{mnp}(A)+
{1\over{2({\partial}_ra)({\partial}^ra)}}
{\partial}^ma{\hat {\cf}}_{mnp}(A){\hat {\cf}}^{npq}(A){\partial}_qa-
{j}^{mn}{A}_{mn}].
\ee
It is invariant under \p{6asymm}--\p{6Asymm2} with the transformation
laws
modified according to \p{mod}. The equations of motion derived from this
action are reduced to the modified self--duality condition
\cite{deser_s,bm}
\be \label{modsd}
{\hat {\cf}}_{mnp}(A)=0;\quad
({\hat {\cf}}_{mnp}(A)={\hat F}_{mnp}(A)-{1\over{3!}}{\eps}_{mnpqrs}
{\hat F}^{qrs}(A)).
\ee
Taking an exterior derivative of \p{modsd} and taking into account
\p{Seq} we obtain the chiral field equations with sources
$$
\partial_m\hat F^{mnp}=j^{np}.
$$

Coupling of matter sources in
the dual formulation based on the
action \p{6dual} is carried out the same way as discussed above, or
can be obtained from \p{6psts}  by the duality transform of the scalar
field.

\section{Conclusion}

We considered duality properties of the Lorentz--covariant
actions for chiral bosons and showed
that the duality transform of the auxiliary scalar field
$a(x)$ produces a dual covariant action which consistently describes
self--dual gauge fields. In the $D=2$ case the formulation considered
turns out to be self--dual with respect to this dualization \cite{mps}.
In
higher dimensions, however, it is not the case and the dual actions
differ
from each other. In $D=4$, for example, as was demonstrated in
\cite{mps},
such a duality transform allows one to connect two different
non--covariant versions of the duality--symmetric Maxwell action
\cite{zwanziger}--\cite{ss}.

We also discussed the problem of coupling the dual actions to external
sources and showed that due to the dyonic nature of the latter the
consistent coupling preserving the symmetry structure of the model
should be non--local (as in formulations considered earlier
\cite{dirac,deser_s,bm}).

Better understanding the origin of the
local symmetries of the duality--symmetric actions and of their
topological structure can provide new information about the quantum
properties of chiral
boson fields and dualities which they are related to.

{\bf Acknowledgements}

We are thankful to P. Pasti, M. Tonin and F. Toppan
for fruitful discussions. One of us (A.M.) is particularly grateful to
the
Organizing Committees of the Dubna Workshop ``Supersymmetry and Integrable
Systems"  and of the Xth School--Seminar ``VOLGA 10'98" for hospitality.
The work of A.M. and D.S. was partially supported by the
INTAS Grants N 93--493--ext and N 97--0308.


\begin{thebibliography}{99}
\bibitem{zwanziger}
D. Zwanziger, \PRD 3 1971 880.
\bibitem{deser}
S. Deser and C. Teitelboim, \PRD 13 1976 1592; \\
S. Deser, {\sl J. Phys.} {\bf A} Math. Gen. {\bf 15} (1982) 1053.
\bibitem{ss}
J.H.  Schwarz and A. Sen, \NPB 411 1994 35.
\bibitem{pst}
P.  Pasti, D.  Sorokin and M. Tonin, \PLB 352 1995 59; \PRD 52 1995
R4277.
\bibitem{pstch1}
P.Pasti, D. Sorokin and M. Tonin, in Leuven Notes in
Mathematical and Theoretical Physics, (Leuven University Press) Series
{\bf B} v.{\bf 6}, (1996) 167.
\bibitem{infinite1}
B. McClain, Y. S. Wu and F.  Yu, \NPB 343 1990 689; \\
C. Wotzasek, \PRL 66 1991 129.
\bibitem{infinite2}
I. Martin and A. Restuccia, \PLB 323 1994 311; \\
F.P. Devecchi and M. Henneaux, \PRD 45 1996 1606; \\
I. Bengtsson and A. Kleppe, {\sl Int. J. Mod. Phys.} {\bf A12} (1997)
3397; \\
N. Berkovits, \PLB 388 1996 743; \PLB 395 1997 28; \PLB 398 1997 79.
\bibitem{pstch2}
P. Pasti, D. Sorokin and M. Tonin, \PRD 55 1997 6292.
\bibitem{fj}
R. Floreanini and R. Jackiw, \PRL 59 1987 1873.
\bibitem{henn}
M. Henneaux and C. Teitelboim, in Proc. Quantum Mechanics of Fundamental
Systems 2, Santiago, 1987, p. 79; \PLB 206 1988 650.
\bibitem{5br}
P. Pasti, D. Sorokin and M. Tonin, \PLB 398 1997 41; \\
I. Bandos, K. Lechner, A. Nurmagambetov, P. Pasti, D. Sorokin and M.
Tonin, \PRL 78 1997 432; \PLB 408 1997 135;\\
M. Aganagic, J. Park, C. Popescu and J. H. Schwarz,
{\sl Nucl.  Phys.} {\bf B496} (1997) 191.
\bibitem{bbs}
I. Bandos, N. Berkovits and D. Sorokin,
{\sl Nucl. Phys.} {\bf B522} (1998) 214.
\bibitem{2bsg}
G. Dall'Agata, K. Lechner and D. Sorokin, \CQG 14 1997 L195; \\
G. Dall'Agata, K. Lechner and M. Tonin, D=10 N=IIB Supergravity:
Lorentz--invariant Actions and Duality, {\bf hep-th/9806140}.
\bibitem{6dn2}
G. Dall'Agata and K. Lechner, \NPB 511 1998 326; \\
G. Dall'Agata, K. Lechner and M. Tonin, \NPB 512 1998 179.
\bibitem{mps}
A. Maznytsia, C. Preitschopf and D. Sorokin, Duality of Self--Dual
Actions, {\bf{hep-th/9805110}}.
\bibitem{si}
W. Siegel, Nucl. Phys. {\bf B238}, 307 (1984);\\
S. J. Gates, Jr. and W. Siegel, Phys. Lett. {\bf B206} (1988) 631.
\bibitem{im}
C. Imbimbo and A. Schwimmer, Phys. Lett. {\bf B193}, 455 (1987).
\bibitem{hull}
C. Hull, Phys. Lett. {\bf B206} (1988) 234; Phys. Lett. {\bf 212B}
(1988) 437.
\bibitem{gates}
D. Depireux, S. J. Gates, Jr. and B. Radak, Phys. Lett. {\bf B236}
(1990) 408.
%\bibitem{cdf}
%C.P. Constantinidis and F.P. Devecchi, \MPLA 13 1998 631;\\
%C.P. Constantinidis, F.P. Devecchi and F. Toppan,
%two--dimensional $N=1,2$ Supersymmetric Chiral and Dual Models,
%{\bf hep-th/9805147}.
\bibitem{cher}
S. Cherkis and J. H. Schwarz,
{\sl Phys. Lett.} {\bf B403} (1997) 225.
\bibitem{deser_s}
S. Deser, A. Gomberoff, M. Henneaux and C. Teitelboim,  \\ \PLB 400 1997
80.
\bibitem{bm}
R. Medina and N. Berkovits, \PRD 56 1997 6388.
\bibitem{dirac}
P. A. M. Dirac, {\sl Proc. R.
Soc.} {\bf A133} (1931) 60, {\sl Phys. Rev.} {\bf 74} (1948) 817.
\end{thebibliography}
\end{document}